\documentclass[12pt]{article}
\usepackage{epsfig}
\usepackage{amssymb,graphicx}
\def\be{\begin{equation}}
\def\ee{\end{equation}}
\def\ba{\begin{array}{c}}
\def\ea{\end{array}}

\newcommand{\bea}{\begin{eqnarray}}
\newcommand{\eea}{\end{eqnarray}}

\newcommand{\kt}{\rangle}

\begin{document}

\begin{center}

{\Large \bf {

Log-anharmonic oscillator and its large$-N$ solution

 }}

\vspace{13mm}

\vspace{3mm}

\begin{center}

\textbf{Miloslav Znojil}\footnote{znojil@ujf.cas.cz}
 and
\textbf{Iveta Semor\'{a}dov\'{a}}\footnote{semoradova@ujf.cas.cz}

\vspace{1mm} Nuclear Physics Institute of the CAS, Hlavn\'{\i} 130,
250 68 \v{R}e\v{z}, Czech Republic

\end{center}

\vspace{3mm}

\end{center}

\subsection*{Keywords:}

.

confining interactions;

soft central repulsion;

large$-N$ expansion method;

logarithmic anharmonicity

\subsection*{PACS number:}
.

PACS 03.65.Ge - Solutions of wave equations: bound states

\section*{Abstract}

Anharmonic oscillator is considered using an unusual, logarithmic
form of the anharmonicity. The model is shown connected
with the more conventional power-law anharmonicity $\sim |x|^\alpha$
in the limit $\alpha \to 0$. An efficient and user-friendly
method of the solution of the model is found in the large$-N$
expansion technique.

\newpage

\section{Introduction\label{Ia}}

The popularity of one-dimensional Schr\"{o}dionger equations with a
wide family of anharmonic-oscillator interactions
 \be
 V^{(AHO)}(x) = \omega^2 x^2 + \lambda \,V_I(x)
 \ee
found its motivation in the phenomenological appeal of the model
(say, in the context of atomic and molecular physics \cite{IJQC27})
as well as in the methodical relevance. Thus, one finds that the
radius of convergence of the most common Rayleigh-Schr\"{o}dinger
perturbation-series representation of the bound-state energies
appears to be zero even after one of the most elementary power-law
choices of perturbations $V_I(x) \sim x^n$ with, typically, quartic
anharmonicity at $n=4$ \cite{Kato}. Such an obstacle forces one to
search for a more sophisticated method of the evaluation and
prediction of the experimental measurements: several nice, compact
and comparatively elementary outlines of the problem may be found,
e.g., in dedicated proceedings \cite{IJQC}.

In the early nineties the study of the one-dimensional anharmonic
oscillators found another specific motivation in the context of
quantum field theory where the choice of $V_I(x) \sim x^\alpha$ with
a non-integer exponent $\alpha \notin \mathbb{N}$ provided an
innovative insight in the mechanism of the spontaneous
symmetry-breaking phenomena \cite{BM} or in the robust nature of
supersymmetry \cite{BM2}. The idea (called, sometimes,
delta-expansion technique \cite{delta}) was based on the truncated
Taylor-series approximation of the general power of the coordinate
or, in the field-theory context, of the field,
 \be
 |\phi|^\alpha= {\rm e}^{\alpha\,\ln |\phi|} =
 1+\alpha\,\ln |\phi| +\frac{1}{2}\,
 \alpha^2\,\ln^2 |\phi| + \ldots\,.
 \label{negl}
 \ee
The latter trick opened a way towards the efficient
perturbation-expansion study of the three-parametric family
 \be
 \left [
 -\frac{d^2}{dx^2} + \omega^2x^2+\lambda\, |x|^\alpha
 \right ]\,\psi_n(x) = E_n\,\psi_n(x)\,,
 \ \ \ \ n = 0, 1, \ldots
 \,
 \label{oklah}
 \ee
of the anharmonic-oscillator Schr\"{o}dinger equations in the three
alternative, phenomenologically different dynamical regimes. In Ref.
\cite{Pade} we studied such a possibility in detail, distinguishing
between the three versions of Eq.~(\ref{oklah}) where (a) $\alpha
\approx 2$, (b) $\alpha \approx 0$ and (c) $\alpha \approx -2$.

The first option (a) may be interpreted as making the spring
constant slightly coordinate-dependent, $\omega^2 \to \omega^2(x) =
\omega^2\,|x|^{\alpha-2}$. Naturally, in all of the three cases (as
well as in their mutual combinations \cite{Pade}) one arrives at the
perturbative Schr\"{o}dinger equation of interest by employing and,
{\it mutatis mutandis}, approximatively truncating the respective
infinite Taylor-series versions (\ref{negl}) of the interaction.

In our present letter our attention will be concentrated upon the
most elementary option (b). In this case one can neglect the ${\cal
O}(\alpha^2)$ corrections, abbreviate $ \lambda\alpha=-2g^2$, shift
the origin of the energy scale and arrive at a remarkably elementary
logarithmically anharmonic Schr\"{o}dinger equation
 \be
 \left [
 -\frac{d^2}{dx^2} + \omega^2x^2-2\,g^2\,\ln |x|
 \right ]\,\psi_n(x) = E_n\,\psi_n(x)\,,
 \ \ \ \ x \in (-\infty,\infty)\,,
 \ \ \ \ n = 0, 1, \ldots\,.
 \label{beklah}
 \ee
Once we arrived at the bound-state problem (\ref{beklah}) we
realized (cf. section \ref{druha} below) that the choice of a small
exponent $\alpha$ in the anharmonicity of Eq.~(\ref{oklah}) (and, in
particular, of its leading-order logarithmic approximation) combines
several features of the interaction endowing the underlying
Schr\"{o}dinger Eq.~(\ref{beklah}) with an independent, purely
mathematical appeal. The most characteristic feature of the
interaction is found to lie in the presence of a central barrier
which is softer than usual. In contrast to the customary centrifugal
barrier $\sim \ell(\ell+1)/x^2$ (which is strictly impenetrable so
that it forces us to restrict the admissible coordinates to a
half-line), the present logarithmic soft barrier $V_I^{(log)}(x)
=-2\,g^2\,\ln |x|$ remains penetrable so that the motion of a
hypothetical particle may proceed along the whole real line of $x
\in (-\infty,\infty)$.

A semi-quantitative analysis of the basic qualitative features and
consequences of the underlying processes of the tunneling through
the logarithmic singularity in the origin has been performed in our
preceding paper \cite{Iveta}. In the generic case we were only able
to construct a purely numerical solution. Moreover, for the sake of
simplicity of this solution we replaced the asymptotically dominant
confining force $\sim \omega^2x^2$ by an infinitely deep square-well
approximation (i.e., after a re-scaling, by the Dirichlet boundary
conditions at $x=\pm 1$).

Near the minimum, such a replacement changed the nature of the force
significantly, especially in the strong-repulsion dynamical regime.
Our present paper will fill the gap. In section \ref{tretia} we
shall return to the untruncated and smooth potential in the
perturbed harmonic oscillator regime. Using the less usual
perturbation-series formalism of the so called large$-N$ expansions
we shall show that a consistent picture of the coupling-dependence
of the bound states becomes provided in the form which may be
characterized as semi-numerical.

In our last section \ref{tretis} we will summarize our message and
emphasize the very satisfactory numerical convergence of our present
large$-N$ results.

\section{The context of quantum mechanics\label{druha}}

\subsection{Special role of the small exponents $\alpha$}

Equation~(\ref{oklah}) represents quantum system, the mathematical
friendliness, probabilistic interpretation and/or physical
applications of which can be different for different parameters
$\omega$, $\lambda$ and $\alpha$. For example, for a rather
artificial choice of negative $\alpha=-2$, the oscillator has to  be
defined, on the half-axis of $x \in (0,\infty)$, as exactly solvable
at any $\lambda>-1/4$ \cite{0102034}. In the context of physics, on
the contrary, the choices of the large and positive $\alpha=4$ or
$\alpha=6$ define the most popular anharmonic oscillators living on
the whole real line of $x \in (-\infty,\infty)$. These oscillators
are important in quantum chemistry \cite{Cizek} as well as in
quantum field theory \cite{speciss}. One of the reasons is that they
remain tractable, at any sufficiently weak coupling $\lambda>0$, by
the standard Rayleigh-Schr\"{o}dinger perturbation theory
(cf.~\cite{Fluegge}, p.~80).

The main qualitative change of the shape of the potential occurs in
the limit $\alpha \to 0$. The sharply spiked shape (and, for
$\lambda>0$, the double-well shape) of the potentials with negative
exponents gets smeared at $\alpha=0$. For the small and positive
$\alpha$ the spike becomes bounded and inessential. Ultimately, it
disappears completely beyond $\alpha=1$. In the vicinity of the
vanishing exponent $\alpha\,$ one may expect the emergence of
phenomena which would depend upon the sign of $\alpha$. The
existence of such a boundary, emphasized in \cite{Pade}, served as
an additional motivation of our interest in Eq.~(\ref{beklah}).

\subsection{Large$-N$ expansions\label{druhab}}

The mathematical essence of large$-N$ method (cf., e.g., its compact
review in Ref.~\cite{Bjerrum} or a small sample of applications in
Refs.~\cite{laN}) can be most easily explained using
Eq.~(\ref{oklah}) in the impenetrable-barrier limit $\alpha \to -2$.
Then, Schr\"{o}dinger equation
 \be
 H(N)\ |\psi_n(N)\kt = E_n(N)\ |\psi(N)\kt\,,
 \ \ \ \ \ n = 0, 1, \ldots
 \label{equati}
 \ee
may be considered with elementary Hamiltonian
 \be
 H(N)=H^{(HO)}(N)=-\frac{d^2}{dx^2} + V^{(HO)}(x)
 \,,
 \ \ \ \
   V^{(HO)}(x)= x^2+\frac{N(N+1)}{x^2}
 \,,
 \ \ \ \ x \in (0,\infty)\,
 \label{impene}
 \ee
and with a {\em large} and real though not necessarily integer $N$.
Our effective potential $V^{(HO)}(x)$ then acquires a unique minimum
at
 \be
 x_{(min)}=R=R(N)=[N\,(N+1)]^{1/4}\gg 1\,.
 \label{fuore}
 \ee
We may replace parameter $N\gg 1$ by its elementary function
(\ref{fuore}). This simplifies the approximation of the potential
near its absolute minimum,
 \be
 V^{(HO)}(R+\xi)=V^{(HO)}(R)+ 4\, \xi^2-\frac{4}{R}\xi^3
 +{\cal O}(\xi^4/R^2)\,.
 \label{trunc}
 \ee
Any higher-precision amendments of this formula may be also
considered. Up to the asymptotically vanishing corrections one
obtains the low-lying spectrum of Eq.~(\ref{equati}) +
(\ref{impene}) in equidistant form
 \be
 E_n^{(HO)}=2R^2+2(2n+1) +{\cal O}(1/R)\,,
 \ \ \ \ n = 0, 1, \ldots\,.
 \label{highero}
 \ee
Within the error bars this result {\em precisely} coincides with the
known exact formula for energy levels $E_n^{(HO)}=4n+2N+3$ where
$n=0, 1, \ldots$.

For some more general potentials or Schr\"{o}dinger equations
(studied, e.g., in Refs.~\cite{three,Omar,jednalomenoel}), the
essence of the success or failure of the whole approach remains the
same. The potential has to develop a deep minimum at a suitable
large parameter (mostly called, in the literature, $N$). Provided
that symbol $R$ denotes the position of the minimum, even the
inspection of our most elementary illustrative example (\ref{trunc})
+ (\ref{highero}) reveals that an amended, more explicatory name of
the technique could read $1/R-$expansion method.

\section{Log-anharmonic oscillator in the large-$N$
approach\label{tretia}}

The potential in Eq.~(\ref{beklah}) is, at small $|x|$, dominated by
an unbounded repulsive spike. In Ref.~\cite{Iveta} we showed that
such a form of the left-right symmetric barrier is soft, i.e., that
it admits tunneling. We restricted attention to the small vicinity
of the origin and we simulated the effects of the asymptotic
confinement using a square-well potential $V_\infty(x)$. Indeed,
such a simplification went partially against the spirit of our
original physical motivation. For this reason we are now returning
to Eq.~(\ref{beklah}) in which the {\it a priori} constraints
imposed upon the potential at large $x$ are smoother, analytic and
more natural.

\subsection{Non-numerical preliminaries}

The overall strategy of our present study of the coexistence of the
logarithmic spike with the harmonic-oscillator asymptotics of the
interaction will be based on two assumptions. Firstly, we shall
accept the results of the comparatively self-contained description
of the system in the weak-coupling setting as given in
Ref.~\cite{Iveta}. Thus, we shall only be interested here in the
strong-coupling dynamical regime in which either $\omega$ is small
or $g$ is large, or both. After a transfer of interest to the
strong-coupling-related phenomena we shall try to clarify the
limitations imposed upon the admissible parameters by the softness
of the barrier.

We will {\em not\,} assume that the applicability of the large$-N$
expansion technique is guaranteed {\it a priori}. In fact, whenever
the central barrier admits a perceivable tunneling one {\em must\,}
expect that such a method of the construction of bound states could
encounter its natural limitation of validity (cf. also paragraph
\ref{paran} below).

Our present verification and confirmation of the applicability of
the large$-N$ expansion techniques to the specific double-well
potential of Eq.~(\ref{beklah}) will have two components. First, we
shall develop an appropriate (i.e., harmonic-oscillator, exactly
solvable, leading-order) approximation of the low-lying-spectrum of
the system in question. Second, a less common test of the method
will be performed. We shall verify (or, depending on the parameters,
disprove) that for the low-lying set of bound states the approximate
harmonic-oscillator potential is a good approximation, i.e., that
its shape more or less coincides with its full-fledged exact
physical predecessor near its minimum.

Although such a test seems to be based on intuitive criteria, it can
also be given a more constructive form via the evaluation of
higher-order corrections. We will check that in certain
strong-coupling dynamical regime the tunneling through the soft
logarithmic barrier is truly suppressed, indeed. We will demonstrate
that for any pre-selected set of the low-lying (and, in principle,
exponentially damped) wave-functions $\psi_n(x)$ their numerical
size becomes negligible {\em before} the coordinate gets any close
to the origin.

In the non-numerical, analytic part of our analysis the shape of our
potential will be taken into account on both of the half-axes. Thus,
say, at $x \in (0,\infty)$ we have
 \be
 V(x)=\omega^2x^2-2g^2 \ln x\,,\ \ \ \ \ \
 V'(x)=2\,\omega^2 x-2g^2/ x\,,\ \ \ \ \ \
 V''(x)=2\,\omega^2+2g^2/ x^2\,,\ \ldots\,.
 \label{billirub}
 \ee
We can easily localize the minimum of the potential at
$x=x_{min}=R=g/\omega $, yielding
 \be
 V(R)=\omega^2R^2-2g^2 \ln R=\left (-2 \ln g+1+2 \ln \omega
 \right )\,g^2
 \,,\ \ \ \ \ \
 V'(R)=0\,,\ \ \ \ \ \
 V''(R)=4\,\omega^2\,,\ \ldots\,.
 \label{extrab}
 \ee
Then, the strong-coupling regime may be characterized by the
property $R=g/\omega \gg 1$. The harmonic-oscillator expansion
(\ref{trunc}) finds its truncated logarithmic-anharmonicity-related
Taylor series analogue in formula
 \be
 V^{}(R+\xi)=V^{}(R)+ 2\, \omega^2\,\xi^2
 +{\cal O}(\xi^3/R)\,.
 \label{newtrunc}
 \ee
This yields our ultimate large$-N$ {\it alias} $1/R-$expansion
prediction
 \be
 E_n=V(R)+\sqrt{2}\,(2n+1)\,\omega +{\cal O}(1/R)\,,
 \ \ \ \ n = 0, 1, \ldots\,
 \label{loghig}
 \ee
of spectrum of the low-lying bound states.


\begin{table}[h]
\caption{A sample of the low-lying energy-level shifts
$\varepsilon_n=E_n-V(R)$ for potential (\ref{billirub}). The
couplings $\omega=0.001$ and $g=1.0$ are the same as in
Fig.~\ref{trija} below. } \label{pexp4}
 \vspace{2mm}
  \centering
\begin{tabular}{||c||c|c|c|c||}
\hline \hline method: & \multicolumn{1}{c}{n=0}&
\multicolumn{1}{c}{n=1}& \multicolumn{1}{c}{n=2}&
\multicolumn{1}{c||}{n=3}\\
\hline \hline large$-N$&
0.00141421& 0.00424264& 0.00707107& 0.00989949\\
\hline numerical& 0.00141432& 0.00424309& 0.00707218& 0.00990161
\\
\hline \hline difference:& -0.00000011& -0.00000045& -0.00000111&
-0.00000212
\\
\hline
 \hline
\end{tabular}
\end{table}

\subsection{Numerical considerations}

As we already indicated, there are two criteria for the
applicability of the approximate spectral formula (\ref{loghig}).
The first condition requires the smallness of $1/R$. This can be
satisfied by our choice of parameters. Beyond this extreme, we could
also localize the boundaries of validity of the approximations via
an explicit perturbative evaluation of corrections or,
alternatively, by the comparison of the approximations with the
results of a suitable numerical method. Table \ref{pexp4} offers a
small sample of such a test.

%
%
%
%
%
%
%
%
%
%
%
%
%
%
%
%


\begin{figure}[h]                    
\begin{center}                         
\epsfig{file=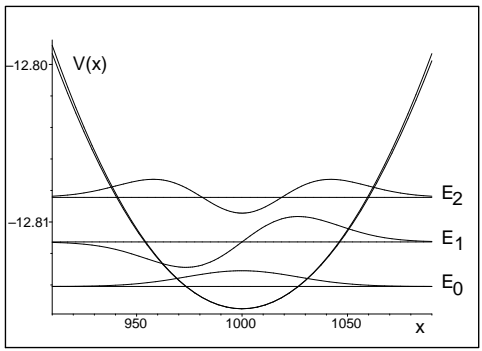,angle=270,width=0.36\textwidth}
\end{center}    
\vspace{2mm} \caption{Logarithmically anharmonic potential
(\ref{billirub}) and its almost identical large$-R$
harmonic-oscillator approximation (\ref{newtrunc}) at
$\omega=1/1000$ and $g=1$. Horizontal lines mark the first three
energy levels. The picture also displays the shapes of the related
wave functions.
 \label{trija}
 }
\end{figure}

The second condition asks for a guarantee of suppression of the
tunneling. This seems to be a purely numerical check. Still, a key
to such a check is analytic. It lies in the availability of the
harmonic-oscillator bound states in closed form. For the ground
state we have
 $$
 \psi_0(x) \sim \exp [-\sqrt{2}\omega\,(x-R)^2/2 ]\,.
 $$
From this formula we can deduce that at a numerical cut-off
$\xi_{max} = R-x_{min}$ the wave function must be already
negligible, i.e., we must have
 $$
 \omega\,\xi_{max}^2 \gg  \sqrt{2}\,.
   $$
It is sufficient to pick up any boundary of the numerically relevant
vicinity of $x=R$,
 $$
 \xi_{max} \gg \sqrt{2}/\sqrt{\omega}
    \,.
 $$
In addition we need that $x_{min}$ remains positive, i.e.,
 \be
 \xi_{max} \ll R\,.
 \label{geneesti}
 \ee
Putting the two constraints together we arrive at our ultimate
safe-approximation constraint
 $$
 g \gg \sqrt{\omega}\,.
 $$
For illustration let us recall Figure \ref{trija} in which we choose
$\omega=0.001$ and $g=1$.  The picture demonstrates that the
difference between the exact and approximate potentials is really
small, more or less comparable with the thickness of the drawing
lines. Secondly, with $R = g/\omega=10^3$, the three lowest-lying
(viz., $n=0$, $n=1$ and $n=2$) bound-state wave functions can be
considered practically vanishing at our tentative choice of
boundaries $\xi_{max}=85$, i.e., already perceivably below our
generic estimate (\ref{geneesti}).

Let us add that just the logarithmic corrections will enter a not
too different asymptotic estimate
 $$
 \psi_n(x) \sim \exp [-\sqrt{2}\omega\,(x-R)^2/2 + {\cal O} (\ln
 |x-R|)]\,,\ \ \ \ \ \ \ |x - R| \gg 1/\sqrt{\omega}
 \,
 $$
valid for the first few excited states. Fig.~\ref{trija} indicates
that the growth of the excitation quantum number $n$ should be
accompanied, in practical calculations, also by a certain not too
quick growth of the safe-estimate value of $ \xi_{max}=
\xi_{max}(n)$.

\subsection{Consequences\label{paran}}

Our preceding considerations have shown that in the strong-coupling
regime with $R \gg 1$ and for the low-lying states of our
log-anharmonic oscillator with the first few quantum numbers $n = 0,
1, \ldots$ the tunneling between the left and right half-axes of $x$
in Eq.~(\ref{beklah}) is suppressed because the wave functions are
very well approximated by their harmonic-oscillator approximants.
These approximants are all exponentially small long before the
coordinate $x$ gets any close to the origin at $x=0$, i.e., they are
exponentially small in the displacement variable $\xi^2={\cal O}
(R^2)$ (cf. Fig. \ref{trija}). The low-lying spectrum of energies
remains almost exactly doubly degenerate.

%
%


\begin{figure}[h]                    
\begin{center}                         
\epsfig{file=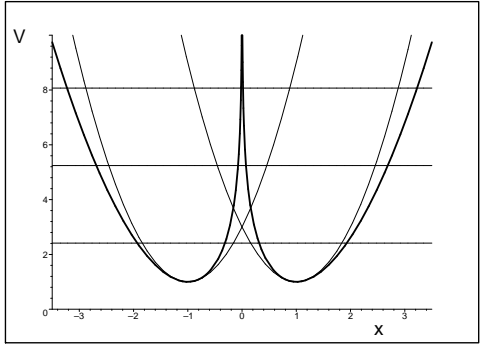,angle=270,width=0.36\textwidth}
\end{center}    
\vspace{2mm} \caption{The emergence of undesirable overlaps of the
two harmonic-oscillator potentials (\ref{newtrunc}) (thin curves),
i.e., a threat of the failure of the large$-N$ approach for the
double well potential (\ref{billirub}) (thick curve) at the
insufficiently large $R=1$ for $\omega=1$ and $g=1$.
 \label{trix}
 }
\end{figure}

In other words, the experimental detection of the effect of the
tunneling would only be possible for the highly excited states or
for the much smaller values of the ratio of couplings $R=g/\omega$.
In particular, the decrease of $R$ would imply the necessity of the
study and incorporation of the higher-order terms in the large$-N$
expansions. Strictly speaking, the technique ceases to be
comfortable and/or sufficiently efficient. Still, the non-numerical
nature of the leading-order harmonic-oscillator large$-N$
approximants preserves its appeal, in spite of the expected
necessity of the inclusion of the higher-order corrections.



\begin{figure}[h]                    
\begin{center}                         
\epsfig{file=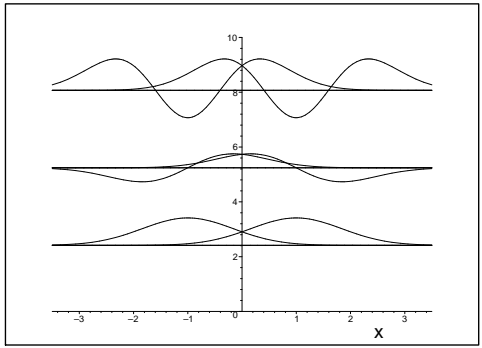,angle=270,width=0.36\textwidth}
\end{center}    
\vspace{2mm} \caption{The emergence of the
significant overlaps of the left and right harmonic-oscillator
wave functions
at
$\omega=1$, $g=1$ and $R=1$ for $n=0$, $n=1$ and $n=3$ indicates that
the degeneracy of the
leading-order horizontal-line
energy levels will break down due to the tunneling.
 \label{trips}
 }
\end{figure}

The emergence of such a scenario is illustrated in Figs.~\ref{trix}
and~\ref{trips}. Non-numerically, the onset of such a necessity may
be characterized by the decrease of the $x=0$ intersection of the
two (viz., left and right) auxiliary potentials (\ref{newtrunc}) at
the height which becomes comparable with the estimated bound-state
energy (viz., close to $E_0$ in Fig.~\ref{trix}).

In the small$-R$ dynamical regime of Fig.~\ref{trix} one encounters
the tunneling effect, i.e., an increase of the overlaps between the
left and right approximate wave functions (cf. Fig.~\ref{trips}).
Their even- and odd-parity linear combinations will become
perceivably different, so that the double degeneracy of the spectrum
will be lost. Still, the split will remain small in the ground state
which feels the thickness of the infinitely high central barrier.



\begin{figure}[h]                    
\begin{center}                         
\epsfig{file=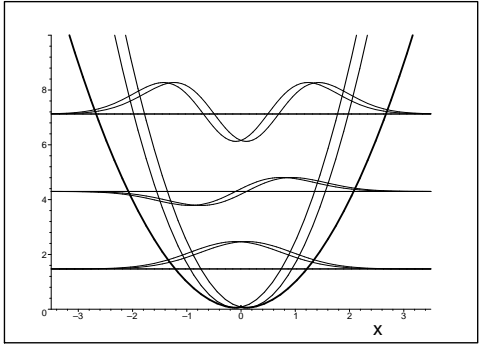,angle=270,width=0.36\textwidth}
\end{center}    
\vspace{2mm} \caption{The disappearance of the
influence of the central
barrier at
$\omega=1$, $g=0.1$ and $R=0.1$.
 \label{trius}
 }
\end{figure}
%
%

A decisive loss of the applicability of the large$-N$ method may be
now illustrated by Fig.~\ref{trius} in which $R=1/10$ is really
small. The analytic large$-N$ estimates will fail, for several
reasons. Firstly, the infinitely high central barrier will become so
thin (in fact, invisible in Fig.~\ref{trius}) that the double-well
structure of the full potential is hardly felt by its wave
functions, exact or approximate. This observation reflects the
``soft'' nature of the central logarithmic spike.

The availability of the approximate energies enables us to see that
the higher excitations will only ``feel'' the logarithmic
singularity perceivably less than their lower-lying predecessors.
Thus, the even-odd-parity degeneracy becomes almost completely
removed.

The superpositions of the higher-order ${\cal O}(1/R)$ corrections
cease to be negligible and seem to act in the single direction.
Indeed, in Fig.~\ref{trius} we clearly see that the left and right
harmonic-oscillator wells prove rather narrow so that the large$-N$
approximation now obviously overestimates the low-lying energy
spectrum in systematic manner.

\section{Concluding remarks\label{tretis}}

Asymptotically confining components $V_\infty(x)$ of
phenomenological potentials are often combined with a short-range
repulsive and impenetrable barrier, say, $V_0(x) =N(N+1)/x^2$. The
large$-N$ expansion techniques are then known to work because the
barrier does not admit any tunneling. A softened, logarithmic
central repulsion $V_0(x) =- N(N+1) \ln |x|\ $ has been considered
here, therefore, as a new and challenging model in which the
tunneling is allowed.

In the context of mathematics we demonstrated that, in principle,
the applicability of the large$-N$ method can survive in multiple
applications of such a type. Their specific feature has been found
in the fact that the precision of the approximation may become
sensitive to the choice of the asymptotic interaction component
$V_\infty(x)$ in practice. This observation is closely correlated
with the underlying motivations in phenomenology. In our preceding
paper \cite{Iveta}, for example, we were interested in the similar
picture of physics along the lines of considerations inspired, e.g.,
by Refs.~\cite{nlseb,nlsec,Kostya}. We studied there a closely
related logarithmic-interaction model which was, incidentally,
non-analytic and, hence, unsuitable for the large$-N$ mathematical
study. Thus, in our present paper we turned attention to an amended,
fully analytic phenomenological model.

The impact of our results upon the understanding of Schr\"{o}dinger
equations with logarithmic singularities might be further enhanced,
in the future, via a replacement of our present logarithmic
interaction term $\sim \ln x$ by its almost equally elementary
power-law-screening modifications $\sim x^{const}\ln x$ and/or by
the more-term superpositions. All of these models will share most of
the technical merits of the elementary $V \sim \ln x$. Typically,
the next-to-trivial one-parametric potential $V_c(x) = x^2 (\ln x -
c)$ will possess the easily-derived Taylor-series representation
near its closed-form double-well minima at $\pm R$ with $ R=
\exp(c-1/2)$. Thus, one can conjecture that the present systematic
large-$N$ approximation technique becomes applicable, without
essential changes, whenever the relevant quantity $R$ appears
sufficiently large.

The smooth nature of our present, most elementary log-anharmonic
model gave us a decisive methodical advantage due to its
non-numerical large$-N$ tractability. This observation is to be read
as the main message delivered by our present paper. It is necessary
to add that the main mathematical part of the task is to be seen in
the details. For example, in the light of the definition of $g^2=-
\lambda\alpha/2$ we imagined that the assumption of a simultaneous
smallness of $\alpha$ and $g^2$ would imply that one stays in the
small-perturbation regime of our preceding paper \cite{Iveta}
whenever the range of $\lambda$ remains bounded.

We turned attention to the genuine strong-coupling dynamical regime
where $\lambda \gg 1$ is large. Naturally, this reopened the
questions of convergence as well as of an appropriate account of the
higher-order corrections in $\alpha$. Indeed, whenever the exponent
$\alpha$ of the general power-law anharmonicity in Eq.~(\ref{oklah})
ceases to be small, one must take into consideration also the
influence of the originally neglected terms in Taylor series
(\ref{negl}). Fortunately, the task is not prohibitively difficult.
The incorporation of the corrections merely leads to a replacement
of our most elementary Schr\"{o}dinger Eq.~(\ref{beklah}) by its
appropriately generalized forms.

For illustration one can consider the single-anharmonicity problem
 \be
 \left [
 -\frac{d^2}{dx^2} + \omega^2x^2-2\,g^2\,(\ln |x|)^p
 \right ]\,\psi_n(x) = E_n\,\psi_n(x)\,,
 \ \ \ \ x \in (-\infty,\infty)\,,
 \ \ \ \ n = 0, 1, \ldots\,
 \label{cekla}
 \ee
containing a new, variable integer $p=1, 2, \ldots$. One finds that
the overall principles of the implementation of the large$-N$ method
at $x>0$ remain unchanged. The $p \neq 1$ update of
Eq.~(\ref{billirub}) reads
 \be
 V(x)=\omega^2x^2-2g^2 (\ln x)^p\,,\ \ \ \ \ \
 V'(x)=2\,\omega^2 x-2pg^2(\ln x)^{p-1}/ x\,,
 \ \ \ \ \ \
 \label{debilli}
 \ee
 $$
 \ \ \ \ \ \
 \ \ \ \ \ \
 \ \ \ \ \ \
 V''(x)=2\,\omega^2-2p(p-1)g^2(\ln x)^{p-2}/ x^2
 +2pg^2(\ln x)^{p-1}/ x^2\,,\ \ \ldots\,.
 $$
Its form implies that the decisive technical step of the
localization of the position of the absolute minimum of the
potential at $x=R$ remains unchanged. Using relation $ V'(R)=0$ we
get
 \be
 R^2=\frac{pg^2}{\omega^2} (\ln R)^{p-1}\,.
 \ee
Whenever $p \neq 1$, this is an implicit and, in general, ambiguous
definition. Still, once we restrict attention to the quantum systems
with a sufficiently large ratio between coupling constants
$g/\omega$ we only have to pick up the maximal root $R$. This
yields, as before, the absolute minimum of the potential in the
strong-coupling dynamical regime (let us leave the necessary lengthy
but straightforward discussion to interested readers).

\subsection*{Acknowledgements}

The project was supported by GA\v{C}R Grant Nr. 16-22945S. Iveta
Semor\'{a}dov\'{a} was also supported by the CTU grant Nr.
SGS16/239/OHK4/3T/14.



\begin{thebibliography}{00}


\bibitem{IJQC27}
J. \v{C}\'{\i}\v{z}ek and E. R. Vrscay, Int. J. Quant. Chem. 21
(1982) 27 - 68.
%

\bibitem{Kato}
T. Kato, Perturbation theory for linear operators. Springer, Berlin, 1966.

\bibitem{IJQC}
P. O. L\"{o}wdin and Y. \"{O}hrn, Eds., Proc. IWPTLO, Int. J. Quant. Chem. 21
(1982) 1 - 214.

\bibitem{BM}
%
C. M. Bender and K. A. Milton, Phys. Rev. D 55 (1997) 3255 - 3259.

\bibitem{BM2}
C. M. Bender and K. A. Milton, Phys. Rev. D 57 (1998) 3595 - 3608.
%
%

\bibitem{delta}
%
C. M. Bender, K. A. Milton, M, Moshe, S. S. Pinsky, and L. M.
Simmons, Jr., Phys. Rev. Lett. 58 (1987)  2615 - 2618.

\bibitem{Pade}
M. Znojil,
Phys. Lett. A 177 (1993) 111 - 20.


\bibitem{Iveta}
M. Znojil and I. Semor\'adov\'a,
Mod. Phys. Lett. A 33 (2018) 1850009.


\bibitem{0102034}
M. Znojil, Phys. Rev. A 61 (2000) 066101.
%


\bibitem{Cizek}
J. D. Louck, J. Mol. Spectrosc. 4 (1960) 298;

V. S. Popov and A. V. Sergeev, Phys. Lett. A 193 (1994) 165;

%
%
D. K. Watson and D. Z. Goodson, Phys. Rev. A 51 (1995) R5.
%


\bibitem{speciss}
%
B. Simon, Int. J. Quant. Chem. 21 (1982) 3 - 26;

T. T. Wu, Int. J. Quant. Chem. 21 (1982) 105 - 118;

A. V. Turbiner and A. G. Ushveridze, J. Math. Phys. 29 (1988) 2053 -
2063.


\bibitem{Fluegge}
S. Fl\"{u}gge, Practical Quantum Mechanics I. Springer, Berlin,
1971.



\bibitem{Bjerrum}
N. E. J. Bjerrum-Bohr, J. Math. Phys.  41  (2000) 2515 - 2536.
%

\bibitem{laN}
U. Sukhatme and T. Imbo, Phys. Rev. D 28 (1983) 418;

L. D. Mlodinow and M. P. Shatz, J. Math. Phys. 25 (1984) 943;

B. Roy, R. Roychoudhury and P. Roy, J. Phys. A 21 (1988) 1579;

F. M. Fern\'{a}ndez, J. Phys. A: Math. Gen. 35 (2002) 10663 - 10667;

H. B\'{\i}la, Czech. J. Phys. 54 (2004) 1049 - 1054;

I. V. Andrianov, V. V.  Danishevskyy and J. Awrejcewicz, J. Sound
Vibr. 283 (2005) 561 - 571;

M. Znojil and U. G\"{u}nther, J. Phys. A: Math. Theor. 40 (2007)
7375 - 7388;
%

A. F. Ferrari, M. Gomes, C. A. Stechhahn,
Phys. Rev. D 82 (2010)
045009;

M. Znojil, Int. J. Theor. Phys. 53 (2014) 2549 - 2557.

\bibitem{three}
M. Znojil,
J. Phys. A: Math. Gen. 36 (2003) 9929 - 9941.

\bibitem{Omar}
O. Mustafa and M. Odeh, J. Phys. A: Math. Gen. 33 (2000) 5207;

M. Znojil, F. Gemperle and O. Mustafa, 
J. Phys. A: Math. Gen. 35 (2002) 5781 - 5793;


O. Mustafa and M. Znojil, 
J. Phys. A: Math. Gen. 35 (2002) 8929 -
8942.

\bibitem{jednalomenoel}
M. Znojil, 
Phys. Lett. A
374 (2010) 807812.


\bibitem{nlseb}
S. De Martino, M. Falanga, C. Godano and G. Lauro, 
Europhys. Lett. 63 (2003) 472.

\bibitem{nlsec}
K. G. Zloshchastiev, 
%
Gravit. Cosmol. 16 (2010) 288 - 297.


\bibitem{Kostya}
K. G. Zloshchastiev and M. Znojil,
Visnyk Dniprop. Univ.
24 (2016) 101 - 107;


M. Znojil, F. R\r{u}\v{z}i\v{c}ka and K. G. Zloshchastiev,
Symmetry 9 (2017) 165.

\end{thebibliography}
\end{document}